\documentclass[twocolumn]{aastex631}

\usepackage{amssymb}
\usepackage{soul}
\usepackage{natbib}
\usepackage{multirow}
\usepackage{graphicx}
\usepackage{tabularx}
\usepackage{verbatim}
\usepackage{color}
\usepackage{ulem}
\usepackage{subfigure}
\usepackage{booktabs,tikz}
\makeatletter
\global\let\tikz@ensure@dollar@catcode=\relax
\makeatother
\usepackage{urwchancal}

\newcommand\soutm{\bgroup\markoverwith
{\textcolor{black}{\rule[0.5ex]{2pt}{0.8pt}}}\ULon}

\submitjournal{ApJ}

\shorttitle{Formation channels of the IGL and ICL}
\shortauthors{Chun et al.}

\graphicspath{{./}{figures/}}

\begin{document}

\title{Formation channels of the diffuse lights in the groups and clusters over time}


\author{Kyungwon Chun}
\affil{Korea Astronomy and Space Science Institute (KASI), 776 Daedeokdae-ro, Yuseong-gu, Daejeon 34055, Korea}
\author{Jihye Shin}
\affil{Korea Astronomy and Space Science Institute (KASI), 776 Daedeokdae-ro, Yuseong-gu, Daejeon 34055, Korea}
\affil{University of Science and Technology (UST), Gajeong-ro, Daejeon 34113, Korea}
\author{Jongwan Ko}
\affil{Korea Astronomy and Space Science Institute (KASI), 776 Daedeokdae-ro, Yuseong-gu, Daejeon 34055, Korea}
\affil{University of Science and Technology (UST), Gajeong-ro, Daejeon 34113, Korea}
\author{Rory Smith}
\affil{Departamento de F{\'i}sica, Universidad T{\'e}cnica Federico Santa Mar{\'i}a, Avenida Vicu{\~n}a Mackenna 3939, San Joaqu{\'i}n, Santiago, Chile}
\author{Jaewon Yoo}
\affil{Quantum Universe Center, Korea Institute for Advanced Study (KIAS), 85 Hoegiro, Dongdaemun-gu, Seoul 02455, Korea}

\begin{abstract}
We explore the formation of the intragroup light (IGL) and intracluster light (ICL), representing diffuse lights within groups and clusters, since $z=1.5$.
For this, we perform multi-resolution cosmological N-body simulations using the ``galaxy replacement technique" (GRT) and identify the progenitors in which the diffuse light stars existed when they fell into the groups or clusters.
Our findings reveal that typical progenitors contributing to diffuse lights enter the host halo with the massive galaxies containing a stellar mass of $10 < \log M_{\rm{gal}}~[M_{\odot}]< 11$, regardless of the mass or dynamical state of the host halos at $z=0$.
In cases where the host halos are dynamically unrelaxed or more massive, diffuse lights from massive progenitors with $\log M_{\rm{gal}}~[M_{\odot}]> 11$ are more prominent, with over half of them already pre-processed before entering the host halo.
Additionally, we find that the main formation mechanism of diffuse lights is the stripping process of satellites, and a substantial fraction (40-45\%) of diffuse light stars is linked to the merger tree of the BCG.
Remarkably, all trends persist for groups and clusters at higher redshifts.
The fraction of diffuse lights in the host halos with a similar mass decreases as the redshift increases, but they are already substantial at $z=1.5$ ($\sim10\%$).
However, it's crucial to acknowledge that detection limits related to the observable radius and faint-end surface brightness may obscure numerous diffuse light stars and even alter the main formation channel of diffuse lights.
\end{abstract}


\section{Introduction}

In the current concordance cosmology of lambda cold dark matter ($\Lambda$CDM), the galaxy cluster, the largest gravitationally bound structure in the universe, grows its mass by merging with many smaller structures.
These smaller structures gravitationally interact with other structures in the cluster and lose their components or even be completely disrupted.
In this process, many stellar components among them spread throughout the cluster.
These diffuse lights are defined as the intracluster light (ICL).

The existence of ICL was first observed by \cite{zwicky1951} in the Coma cluster, and many studies have demonstrated that the ICL is a ubiquitous structure in many clusters \citep[e.g.,][]{mihos2005,janowiecki2010,rudick2010,iodice2016,montes2018,jimenez-teja2019,spavone2020,yoo2021}.
Moreover, the diffuse lights also manifest as intragroup light (IGL) in group-sized halos \citep[e.g.,][]{darocha2005,darocha2008,raj2020,poliakov2021,ragusa2021}.

The formation of diffuse lights through tidal interaction in the hierarchical structure formation process suggests that their properties offer valuable insights into the evolution of both groups and clusters.
Furthermore, recent studies have shown that diffuse lights can serve as tracers for the distribution of dark matter, determining the shape of potential well in the host halo \citep[e.g.,][]{montes2019,asensio2020,sampaio2021,yoo2022,shin2022}.

Many observational and theoretical studies have proposed several formation channels for diffuse lights, including in-situ star formation \citep[e.g.,][]{puchwein2010}, tidal disruption of dwarf galaxies \citep[e.g.,][]{janowiecki2010}, tidal stripping of massive galaxies \citep[e.g.,][]{rudick2009,puchwein2010,montes2014,contini2018,ragusa2021}, major mergers of satellite galaxies with the Brightest Cluster Galaxy (BCG) \citep[e.g.,][]{murante2007,contini2018}, and pre-processing in group-sized halos \citep[e.g.,][]{rudick2006}.
While each channel contributes to the formation of diffuse lights, their relative importance varies with the evolutionary stages of the clusters \citep[e.g.,][]{ko2018}.

Although diffuse lights are ubiquitous structures in the evolved massive structures, there is considerable variation in the fraction of the diffuse lights to the total luminosity in the structures as 2-40\% \citep{burke2015,morishita2017,jimenez-teja2018,montes2018,spavone2018,furnell2021,yoo2021,poliakov2021,ragusa2021,joo2023}.
This variability may stem from the unique properties of host halos, but the low surface brightness of diffuse lights can make it difficult to measure their fraction accurately. 
Additionally, observing diffuse lights at higher redshifts is particularly challenging due to the cosmological dimming effect.
Despite this difficulty, observation studies have revealed the presence of diffuse lights beyond $z=1$ \citep[e.g.,][]{adami2013,ko2018,coogan2023,joo2023,werner2023}.
In particular, \cite{joo2023} found that the ICL is already abundant ($\sim17\%$) within ten clusters at $1 < z < 2$.
The presence of the abundant ICL in the high-z clusters indicates the necessity of extended studies on the formation of the diffuse lights beyond $z=1$.

Recent simulation studies \citep{contini2023,werner2023} show results consistent with observations by \cite{joo2023}.
However, \cite{contini2023} proposed stellar stripping as the main formation channel of the ICL in the simulations using a semianalytical model, whereas \cite{werner2023} advocated for pre-processed ICL originating from accreted halos in Hydrangea hydrodynamical simulations \citep{bahe2017}.
This difference can be biased by the properties of the simulated halos, but it is possible that the differences in the simulation techniques and/or the ICL definition influence it.

Since the absence of a unique ICL definition, previous studies have used various detection methods, including surface brightness limit \citep[e.g.,][]{rudick2011,burke2015}, surface brightness profile \citep[e.g.,][]{gonzalez2007,goldenmarx2023,joo2023}, or wavelet-based method \citep[e.g.,][]{darocha2005,adami2013,jimenez-teja2016,ellien2021}. 
Different definitions can change the ICL fraction even within the same clusters \citep[e.g.,][]{rudick2011,presotto2014,cooper2015,montes2018,kluge2021,chun2022}.
For instance, \citet{montes2018} showed a significant difference in the ICL fractions in six massive observed clusters when applying two different methods (surface brightness limit and surface brightness profile).
Similarly, \citet{rudick2011}, through controlled N-body simulations, demonstrated that the ICL fraction of clusters could differ by up to a factor of three based on the three different detection definitions (boundness, 3D density limit, and surface brightness limit).
Moreover, the ICL fraction is influenced by the detection limits, such as the choice of observable radius determined by the telescope's Field-of-View (FoV) and faint-end surface brightness limit \citep{burke2015,montes2018}.
Thus, a comprehensive statistical analysis of the ICL requires a large cluster sample, a standardized ICL definition, and accurate quantification of detection limits.

In this work, we investigate how the formation channels of diffuse lights and their fraction within the groups and clusters change since $z=1.5$.
Additionally, we aim to understand how detection limits related to the observable radius and faint-end surface brightness limit influence the properties of diffuse lights.
To achieve this, we perform multi-resolution cosmological N-body simulations using the ``galaxy replacement technique" (GRT) introduced first in \cite{chun2022} for 84 host halos of $13.6 < \log M_{\rm{200c}} < 14.4$, where $M_{\rm{200c}}$ is the mass contained within a radius where the density equals 200 times the critical density of the universe at a specific redshift. 
The GRT allows us to trace the spatial distribution and evolution of massive structures and their substructures without including computationally expensive baryonic physics.
While a halo with $M_{\rm{200c}} < 10^{14}~M_{\odot}$ is conventionally categorized as a galaxy group, for the sake of convenience, we refer to all host halos as galaxy clusters in this study.
Therefore, the diffuse lights in all host halos are called the ICL.

This paper is structured as follows:  In Section \ref{sec:simulation}, we introduce the GRT method and define the ICL within simulated clusters.
Section \ref{sec:dstate} explores the formation channels of the ICL depending on the mass and dynamical states of the clusters at $z=0$.
In Section \ref{sec:evolution}, we delve into the formation and evolution of the ICL since $z=1.5$.
Lastly, we discuss and summarize our results in Section \ref{sec:discussion} and \ref{sec:summary}.

\section{Simulation} 
\label{sec:simulation}

\subsection{Galaxy replacement technique} 
\label{sec:GRT}
In this work, we examine the evolution of the ICL in 84 clusters using the GRT, which was first introduced in \cite{chun2022}. 
The GRT does not include hydrodynamical recipes (e.g., radiative cooling/heating, star formation, supernova feedback, etc.) that require many computational resources.
Instead, we insert high-resolution models of galaxies, including a dark matter halo and stellar disk, into cosmological N-body simulations to study the impact of gravitational tides in a fully cosmological context. 
It enables us to trace the spatial distribution and evolution of the ICL in a short computational time.
Thus, we can achieve enhanced mass and spatial resolution to model the tidal stripping process and the formation of very low surface brightness features.
To perform a simulation with the GRT, we first perform the DM-only cosmological simulation as the base simulation.
For this, we use a set of 64 DM-only cosmological simulations of a (120$~$Mpc $h^{-1}$)$^3$ uniform box with 512$^3$ particles using the cosmological simulation code Gadget-3 \citep{springel2005}. 
This simulation set is called the N-cluster run and is used in many works \citep{smith2022a,smith2022b,chun2022,jhee2022,kim2022,yoo2022,awad2023,chun2023,dong2024}.
In the N-cluster run, the DM particle mass is $~10^9~M_{\odot} h^{-1}$, and the gravitational softening length is fixed at 2.3$~$kpc $h^{-1}$ on a comoving scale.
The lower mass limit of a halo is $~2\times10^{10} M_{\odot} h^{-1}$ ($N_{DM} = 20$).
Among the halos at $z=0$, we select 84 clusters ($13.6 < log M_{200c} [M_{\odot}] < 14.8$) and refer to these clusters as the ``GRT clusters".

To trace the evolution of the GRT clusters with higher-resolution particles, we substitute all low-resolution DM halos that will later constitute the growth of the GRT clusters with a high-resolution galaxy model.
As halos with $M_{\rm{peak}} < 10^{11}~M_{\odot}~h^{-1}$ consist of fewer than 100 low-resolution DM particles and are expected to contribute less than 2\% of the total stellar mass in the clusters at $z=0$, we only replace halos with $M_{\rm{peak}} > 10^{11}~M_{\odot}~h^{-1}$, where $M_{\rm{peak}}$ is the maximum mass of a DM halo before falling into a more massive halo.
To consider a reasonable stellar mass and the build-up of realistic stellar structures in the halos by removing the inflow of the low-resolution DM particles, which might result in artificial heating effects, each DM halo is replaced with the high-resolution galaxy model when one of the following criteria is first satisfied: (1) the halo reaches $M_{\rm{peak}}$, (2) the halo first accretes a replaced satellite. 

Finally, we perform the multi-resolution resimulations to trace the evolution of the GRT clusters until $z=0$ \footnote{In the N-cluster run and the GRT simulations, the post-Planck cosmological model of $\Omega_{m} = 0.3,~\Omega_{\Lambda} = 0.7,~\Omega_{b} = 0.047$, and $h = 0.684$ is used.}.
The high-resolution particle mass for DM and star is 5.4$\times$10$^{6}~M_{\odot}~h^{-1}$ and 5.4$\times$10$^{4}~M_{\odot}~h^{-1}$ and their gravitational softening length is $\sim$ 100 and $\sim$ 10 pc$~h^{-1}$, respectively.
The mass and softening length of low-resolution DM particles are $~10^9~M_{\odot} h^{-1}$ and 2.3$~$kpc $h^{-1}$, respectively, the same as those of the low-resolution DM-only simulation.

\begin{figure*}
\centering
\includegraphics[width=0.95\textwidth]{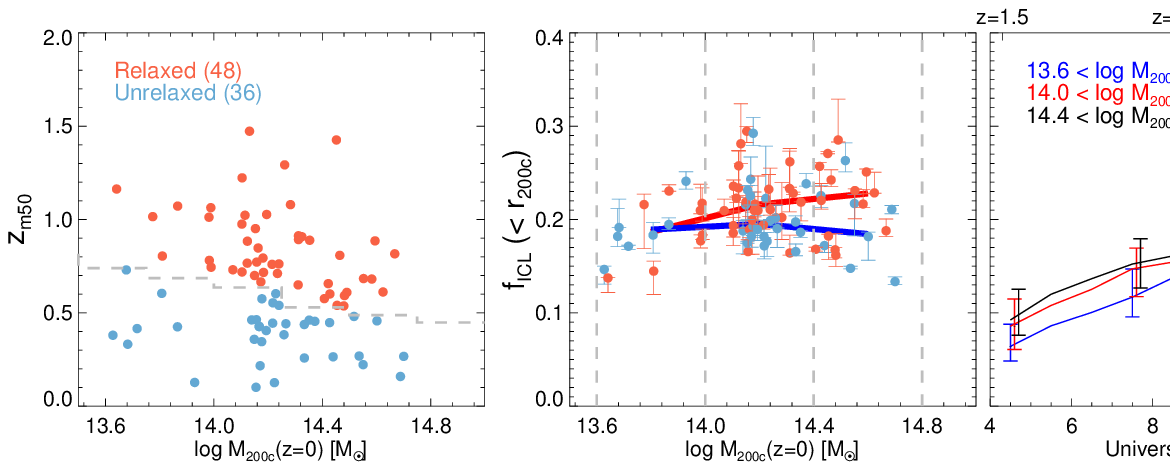}
\caption{\textbf{Left panel:} the relation between $M_{\rm{200c}}$ and $z_{\rm{m50}}$ of the GRT clusters at $z=0$. The filled red and blue circles indicate the relaxed and unrelaxed clusters. The dashed gray line is the median $z_{\rm{m50}}$ of all clusters in the N-cluster run within a logarithmic mass bin of 0.25 dex. \textbf{Middle panel:} the relation between $M_{\rm{200c}}$ and ICL fraction ($f_{\rm{ICL}}$) within $r_{\rm{200c}}$ of the GRT clusters at $z=0$. The filled red and blue circles indicate the median $f_{\rm{ICL}} (< r_{\rm{200c}})$ among the ICL fractions calculated on the three different planes for relaxed and unrelaxed clusters. The error bar caps are the upper and lower values of the ICL fractions for each cluster. \textbf{Right panel:} the evolution of the ICL fraction within the $r_{\rm{200c}}$ of the GRT clusters at $z=0$ since $z=1.5$, depending on the mass of host clusters. The solid blue, red, and black lines indicate the evolution of the GRT clusters within the three host mass bins of $13.6 < \log M_{\rm{200c}} < 14$, $14.0 < \log M_{\rm{200c}} < 14.4$, and $14.4 < \log M_{\rm{200c}} < 14.8$ at $z=0$, respectively. The error bar caps indicate the first and third quartiles of the ICL fractions at each redshift.}
\label{fig:ficl}
\end{figure*}

The left panel of Figure \ref{fig:ficl} shows the mass distribution of 84 GRT clusters and their $z_{\rm{m50}}$ parameter at $z=0$.
The $z_{\rm{m50}}$ parameter is defined as the epoch when the clusters first acquire half of their virial mass.
We use the $z_{\rm{m50}}$ parameter to quantify the mass-growth history of each cluster and to represent the dynamical state of each cluster.
Indeed, previous simulation studies have shown that the clusters with higher $z_{\rm{m50}}$ are more dynamically relaxed \citep[e.g.,][]{power2012,mostoghiu2019,haggar2020,chun2023}.
In this work, to classify the relaxed and unrelaxed clusters among the GRT clusters, we compute the median value of $z_{\rm{m50}}$ within a logarithmic mass bin of 0.25 dex using all clusters within a mass range of $M_{\rm{200c}}=10^{13-15}~M_{\odot}$ in the N-cluster run (the dashed gray line).
We classify the clusters with $z_{\rm{m50}}$ higher than the median value of $z_{\rm{m50}}$ as the relaxed clusters (the filled red circles) and others as the unrelaxed clusters (the filled blue circles).
This panel shows that the $z_{\rm{m50}}$ parameter decreases as the mass of clusters increases due to the hierarchical mass growth in $\Lambda$CDM cosmology.
Note that this classification is only valid for the clusters at $z=0$ because it uses the properties of the clusters at $z=0$.
Therefore, the formation channels of the ICL in the GRT clusters according to their dynamical state are only investigated at $z=0$ in Section \ref{sec:dstate}.

\subsection{ICL detection} 
\label{sec:definition}

In this work, we use the surface brightness limit (SBL) to define the ICL.
Although the SBL for the ICL may lack direct physical significance, it is worth noting that this method is convenient for directly comparing the properties of the ICL in observations with those in simulations \citep[e.g.,][]{rudick2006,rudick2011,cui2014,tang2018,asensio2020}.
Furthermore, \cite{chun2022} demonstrated that the three different ICL detection methods, defined by the SBL, 3D density, and boundness, show a similar temporal evolution of the ICL and BCG fraction.
\cite{rudick2011} also showed similar results on the ICL.
The similarity of the temporal evolution in the three detection methods suggests that the methods respond similarly to events such as mergers and satellite disruptions. 
Note that because we define the ICL using the SBL, the stars bound to the outer halo of a galaxy but fainter than the SBL are also classified as ICL stars.

Although there is no unique SBL of the ICL, we use a V-band surface brightness of $\mu_{V}=26.5~\rm{mag}~\rm{arcsec^{-2}}$, which is the value many works have used \citep{rudick2006,rudick2011,cui2014,presotto2014,tang2018,asensio2020}.
In this scheme, the ICL components are defined as the stellar components in regions fainter than the SBL.
Naturally, the others are classified as the stellar components of the galaxies.
We additionally define a galaxy in the most central region as the BCG.
Due to the resolution of the GRT simulation, we can resolve the stellar structures brighter than the faint-end SBL of $\mu_{V}=31~\rm{mag}~\rm{arcsec^{-2}}$.

To generate the surface brightness map of the GRT clusters, we utilize the projected 2D distribution of stellar particles within the $R_{\rm{200c}}$ of each cluster. 
The stellar distribution is projected onto three different planes: the x-y, x-z, and y-z planes.
Subsequently, the stellar particles are binned into a grid of cells with a length of $D\sim1~$ckpc\footnote{This length corresponds to the physical scale of a CCD pixel ($\sim 2\arcsec$) if clusters are 100~Mpc distant.
This particular CCD pixel specification is the same as that of the K-DRIFT (KASI-Deep Rolling Imaging Fast-optics Telescope) pathfinder, which is a new telescope optimized for LSB studies \citep{byun2022}.}.
This process allows us to investigate three different ICL distributions within each cluster depending on the projected planes.

Because we perform collisionless N-body simulations in this work, we assign the formation epoch and metallicity to each stellar particle in all replaced halos using the following assumptions: (1) Formation epoch: To determine the formation epoch of each stellar particle, we use the mass growth history of each replaced halo and the stellar-to-halo mass relation of \cite{behroozi2013a}. This allows us to estimate the stellar mass growth history of the halo up to the time when it is replaced by the high-resolution model. We then calculate the number of newly formed stars in the halo at each snapshot. The formation epoch of these newly formed stars is determined as the redshift at each snapshot. (2) Metallicity: We estimate the metallicity of the newly formed stars at each snapshot based on the total stellar mass within the halo. To achieve this, we employ the galaxy mass-metallicity relation of \cite{ma2016}.
Lastly, when the halo is replaced with the high-resolution model, we assume that the stellar particles in the inner region of the stellar disk have a higher metallicity and a lower formation epoch.
The temporal evolution of the luminosity of each stellar particle is calculated by using the stellar population synthesis model of \cite{bruzual2003}.

The middle panel of Figure \ref{fig:ficl} shows the ICL fraction of the GRT clusters at $z=0$.
The ICL fraction ($f_{\rm{ICL}}$) is defined as the fraction of the ICL luminosity to the total stellar luminosity within $r_{\rm{200c}}$ of the cluster. 
In the figure, the filled circle with an error bar represents a scatter of three different ICL fractions for each cluster obtained from the stellar distribution projected onto three different planes. 
In order words, the filled circle indicates the median ICL fraction for each cluster, while the upper and lower caps of the error bar represent the maximum and minimum values of the ICL fractions for each cluster.
Despite the diversity (10-30\%) in ICL fractions among the GRT clusters, we find that the median ICL fraction for relaxed clusters (the solid red line) is higher compared to unrelaxed clusters (the solid blue line).
This difference shows the possibility that the dynamical state of clusters can play a role in the ICL fraction of clusters, as suggested by observation studies \citep[e.g.,][]{darocha2008,montes2018,ragusa2021,ragusa2023}.

Many previous simulation studies have shown that, during the growth of the cluster, the ICL fraction of a cluster continuously increases with time, as numerous galaxies fall into the cluster and some of their stars escape from the galaxies \citep[e.g.,][]{rudick2006,rudick2009,puchwein2010,rudick2011,contini2014,harris2017,contini2018}.
This evolutionary trend is also shown in the GRT clusters (The right panel of Figure \ref{fig:ficl}).
In the figure, the solid blue, red, and black lines indicate the evolution of the median ICL fraction of the clusters within the three host mass bins of $13.6 < \log M_{\rm{200c}} < 14$, $14.0 < \log M_{\rm{200c}} < 14.4$, and $14.4 < \log M_{\rm{200c}} < 14.8$ at $z=0$, respectively. 
The error bars denote the range between the first and third quartiles of $f_{\rm{ICL}}$ at some redshifts.
We can see that the ICL fractions of the clusters are increasing steadily from $z\sim1.5$, regardless of the mass of clusters at $z=0$.
However, the fractions remain roughly constant after 10$\sim$11~Gyr in the age of the universe. 
Moreover, we can see that the ICL fractions of the clusters do not significantly increase with the mass of clusters, as suggested by previous observational studies \citep[e.g.,][]{furnell2021,montes2022,ragusa2023}.

\section{Formation channels of the ICL at \lowercase{$z=0$}} 
\label{sec:dstate}

Individual clusters have unique mass growth histories, and therefore, they can be found today in a wide variety of dynamical stages.
The different evolution histories lead to differences in the properties of clusters and subhalos they accrete \citep[e.g.,][]{decarvalho2019,morell2020,chun2023,park2023}.

\begin{figure}
\centering
\includegraphics[width=0.49\textwidth]{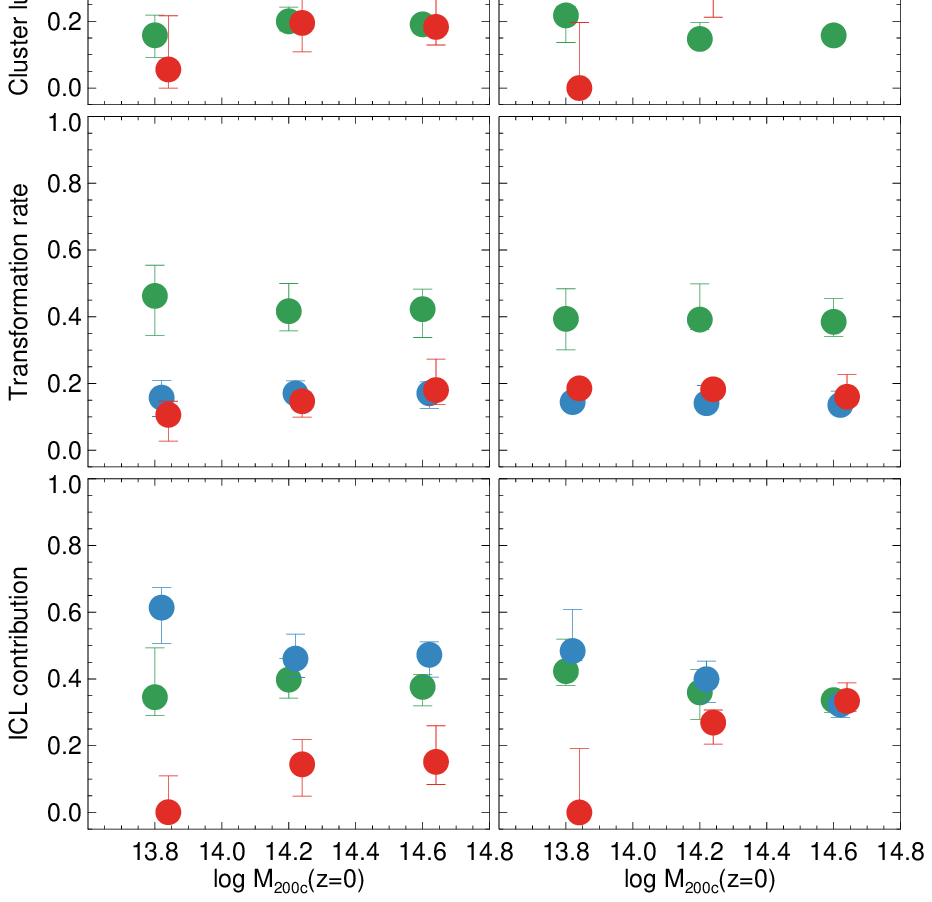} 
\caption{The contribution to the luminosity within the relaxed and unrelaxed GRT clusters depending on the mass of satellites. From top to bottom, each panel indicates, from the subgroups divided depending on the mass of the satellites, the contribution to the total luminosity within the clusters, the fraction entering the ICL regime, and the contribution to the total ICL luminosity. The filled red, blue, and green circles indicate the subgroups called the ``high-M infaller", ``intermediate-M infaller", and ``low-M infaller", and each error bar cap means the first and third quartiles for each subgroup. The meaning of each subgroup is described in Section \ref{sec:dstate}.}
\label{fig:fstar_z0}
\end{figure}

The top panels of Figure \ref{fig:fstar_z0} show how the accreted satellites contribute to the total luminosity of the clusters.
We categorize the accreted satellites into three distinct subgroups using their galaxy stellar mass at the infall time: the ``low-M infaller" (the filled green circles; $M_{\rm{gal}} < 10^{10} M_{\odot}$), ``intermediate-M infaller" (the filled blue circles; $10^{10} M_{\odot} < M_{\rm{gal}} < 10^{11} M_{\odot}$), and ``high-M infaller" (the filled red circles; $M_{\rm{gal}} > 10^{11} M_{\odot}$).
Here, the $M_{\rm{gal}}$ is the combined stellar mass of their central galaxy and diffuse light at the infall time ($z_{\rm{infall}}$).
Thus, the substructures of each accreted satellite are classified into each subgroup, independent of the host satellite, according to their own $M_{\rm{gal}}$.
As we do not consider the in-situ star formation within the cluster, the total mass of stellar components indicates the sum of $M_{\rm{gal}}$ of accreted satellites.
Each symbol indicates the median contribution of accreted satellites to the total luminosity of the clusters, categorized into the three host mass bins shown in the legend of Figure \ref{fig:ficl}. 
The error bars denote the range between the first and third quartiles, providing a distribution of sample clusters.

In the top panels, we can see that the contribution of ``intermediate-M infallers" predominates in both relaxed and unrelaxed clusters, regardless of the mass of clusters.
In contrast, the contribution of ``high-M infallers" escalates with increasing the mass of clusters.
This trend is more clearly visible in the unrelaxed clusters. 
This is attributed to the fact that the unrelaxed clusters tend to undergo more recent mass accretion events than their relaxed counterparts, and thus, they frequently experience merging events with more massive structures.
Unlike the significant trend of two different infallers, the contribution of ``low-M infallers" remains relatively consistent ($\sim$20\%), regardless of the mass and dynamical state of clusters.

As these satellites gravitationally interact with other structures in the cluster, they lose some parts of stellar components or can even be completely disrupted. 
This transformation of satellites naturally generates the ICL within the cluster \citep[e.g.,][]{zibetti2005,conroy2007,murante2007,burke2015}.
The middle panels show the transformation rate at which stars in the satellites enter the ICL regime in the relaxed and unrelaxed clusters.
The panels show that 40\% of the stars in ``low-M infallers" enter the ICL regime, while the transformation rate of more massive satellites is relatively low (10-20\%).
This is because the ``low-M infallers" prefer to contribute to the ICL rather than the BCG, defined as the galaxy in the most central region of the cluster.
Indeed, 68-85\% of both the ICL and BCG stars contributed by the ``low-M infallers" are categorized as ICL stars at $z=0$.
These fractions are sufficiently higher than those of the ``intermediate-M infallers" and ``high-M infallers" (20-48\%).

As the transformation rate of ``low-M infallers" to the ICL is higher than others, we can expect their significant contribution to the ICL stars at $z=0$, although their contribution to the total luminosity of clusters is relatively minimal, as shown in the top panels of Figure \ref{fig:fstar_z0}.
Indeed, the bottom panels of Figure \ref{fig:fstar_z0} show that the ``low-M infallers" contribute significantly to the ICL stars at $z=0$, 30-40\% of $f_{\rm{ICL}}$ ($z=0$).
In contrast, the fraction of stars from the ``intermediate-M infallers" that transform into ICL stars is small, but they still constitute a noteworthy portion of ICL because of their substantial portion of the total luminosity in the clusters.
Unlike these satellites, which contribute significantly to the formation of the ICL at $z=0$, irrespective of the mass and dynamical state of clusters, the ``high-M infallers" have a more significant contribution to the ICL in more massive and unrelaxed clusters.
The importance of group infall in forming the ICL in the dynamical young clusters has indeed been observed in a Virgo cluster \citep{mihos2005,mihos2017}.

\begin{figure}
\centering
\includegraphics[width=0.49\textwidth]{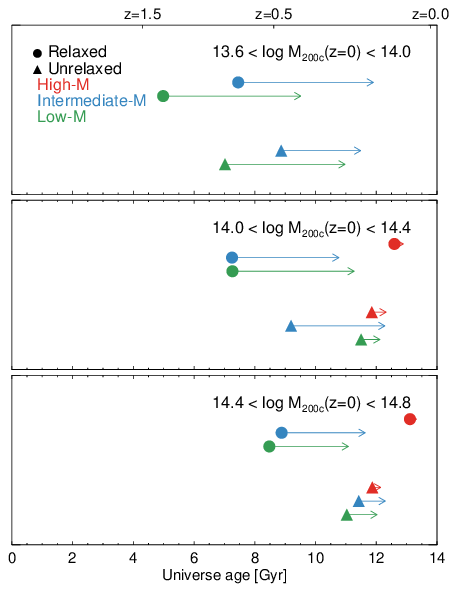}
\caption{The epoch of accretion and transformation of the ICL progenitors in the GRT clusters at $z=0$. The filled symbols and arrowheads indicate when the 80\% of ICL stars at $z=0$ were accreted on the clusters and entered the ICL regime within the clusters. The filled circles and triangles mean the relaxed and unrelaxed clusters, and each color of symbols indicates the subgroups depending on the mass of satellites, shown as Figure \ref{fig:fstar_z0}.}
\label{fig:ficl_z0}
\end{figure}

The unique merging history of each cluster suggests that the ICL progenitors may have different accretion times, dependent on the properties of the clusters \citep[see][]{chun2023}.
Figure \ref{fig:ficl_z0} presents when 80\% of ICL stars at $z=0$ were accreted on the host cluster and transformed into the ICL components.
Here, the epoch of accretion and transformation of stars means the time when their progenitor initially entered the virial radius of the host cluster and when they first entered the ICL regime ($\mu_{\rm{V}}>26.5~\rm{mag}~\rm{arcsec^{-2}}$) as they spread from a main stellar body of the progenitor, respectively.
The accretion time of satellites in the relaxed and unrelaxed clusters is represented by the filled circles and triangles, respectively, and each arrowhead indicates the transformation time.
Thus, the size of the arrows represents the time interval during which stars enter the ICL regime after the accretion of their progenitor.

In each panel, we can see that the ``low-M infallers" and ``intermediate-M infallers" contributing to the ICL star (the filled green and blue symbols) entered into the relaxed clusters earlier compared to their counterparts in the unrelaxed clusters due to the earlier growth of relaxed clusters. 
Furthermore, since more massive clusters can actively accrete the satellites until more recently compared to less massive clusters with the same dynamical state (see the left panel of Figure \ref{fig:ficl}), the accretion of ``low-M infallers" and ``intermediate-M infallers" contributing to the ICL stars occurs more recently.

Within the cluster environment, the stars can be removed from the main body of their host progenitor by the various gravitational interactions \citep[e.g.,][]{rudick2009,smith2016,bahe2019}, and some of them enter into the ICL regime.
Each panel shows that the ICL stars originating from the “low-M infallers” and “intermediate-M infallers” in the relaxed or less massive clusters have had sufficient time to enter the ICL regime as they entered into the host cluster earlier than their counterparts in the unrelaxed or massive clusters.
On the other hand, the stars from ``high-M infallers" rapidly enter the ICL regime after the accretion of their progenitors, regardless of the mass and dynamical state of the host cluster.
This accelerated transformation is attributed to the presence of pre-processed ICL stars.

\begin{figure}
\centering
\includegraphics[width=0.49\textwidth]{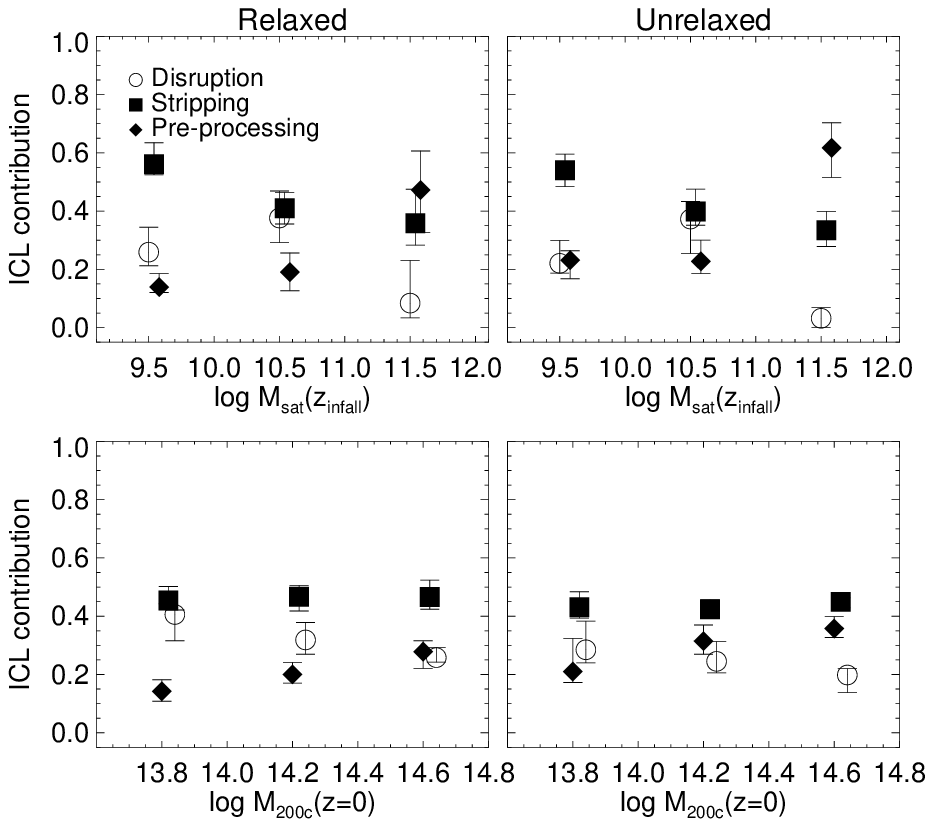}
\caption{The relation between the mass of satellites at $z_{\rm{infall}}$ (upper panels) or the mass of clusters at $z=0$ (lower panels) and the contribution of satellites to the ICL within the clusters, categorized depending on the three different formation mechanisms. The open circles, filled squares, and filled diamonds indicate the contribution of disrupted ICL, stripped ICL, and pre-processed ICL to the total ICL stars.}
\label{fig:type_z0}
\end{figure}

Figure \ref{fig:type_z0} shows the contribution of satellites to the ICL stars at $z=0$, categorized according to the ICL formation mechanisms: the disruption, stripping, and pre-processing.
In this work, the stripped (or disrupted) ICL stars are classified as stars that entered the ICL regime before (or after) their host progenitor was disrupted within the cluster.
Here, we assume that the progenitor is disrupted when its mass reaches the lower mass limit of the N-cluster run, approximately $2\times10^{10} M_{\odot} h^{-1}$, or it is not found anymore.
It is crucial to note that, due to the definition of the ICL in this study, the stripped ICL stars include not only stars entirely removed from satellites but also stars fainter than $\mu_{V}=26.5~\rm{mag}~\rm{arcsec^{-2}}$ bound to satellites.
On the other hand, the ICL stars are classified as pre-processed ICL stars if they have already entered the ICL regime before their progenitor first crossed the virial radius of the host cluster.
Note that this allows the presence of the pre-processed ICL not only in the ``high-M infallers" but also in the ``low-M infallers" and ``intermediate-M infallers".

The upper panels indicate which ICL formation mechanism is preferred by the satellites depending on their galaxy stellar mass. 
In these panels, although we do not separate the contribution of satellites according to the mass of host clusters, we find the trend is not changed by the mass of host clusters.
As mentioned above, in the ``high-M infallers", the pre-processing is the dominant mechanism for the ICL formation.
On the other hand, the ICL formation by the stars from the ``low-M infallers" and ``intermediate-M infallers" occurs within the cluster environment.
In particular, the stripping is the most preferred mechanism in the case of these satellites.

Because most ICL stars at $z=0$ came from the ``low-M infallers" and ``intermediate-M infallers", the stripped ICL stars dominantly contribute to the ICL stars at $z=0$ (40-50\%), regardless of the mass and dynamical state of the clusters (the lower panels of Figure \ref{fig:type_z0}). 
Unlike the relatively constant contribution of the stripped ICL within a wide mass range of clusters, the contribution of the pre-processed ICL to the total ICL stars at $z=0$ tends to be higher in more massive clusters or in the dynamically unrelaxed clusters.
This trend is similar to the contribution of the ``high-M infallers" to the ICL, as more than half of ICL stars from these infallers are classified as pre-processed ICL.

\section{Formation and evolution of the ICL} 
\label{sec:evolution}

\subsection{Formation channels of the ICL since \lowercase{$z=1.5$}}
\label{sec:evolution1}

\begin{figure*}
\centering
\includegraphics[width=0.95\textwidth]{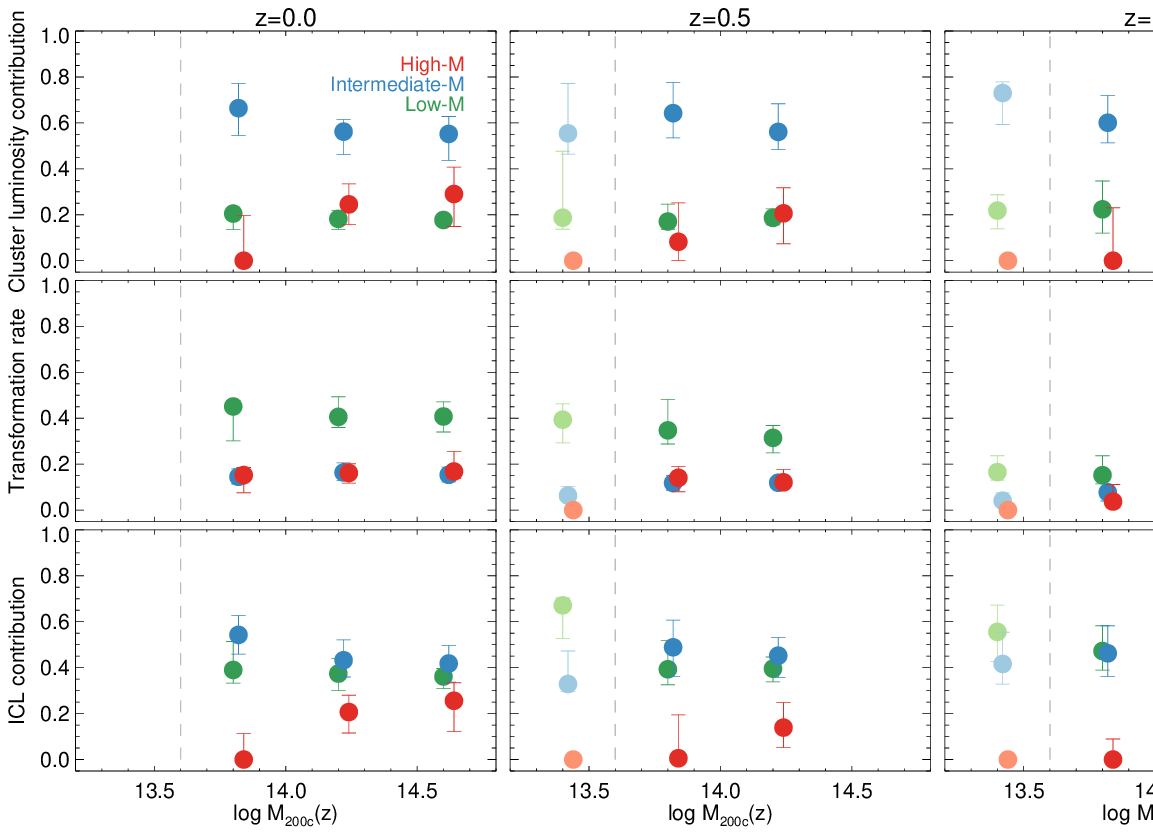}
\caption{The contribution to the luminosity within the GRT clusters at $z=0$, $z=0.5$, and $z=1.5$ depending on the mass of satellites. From top to bottom, each panel indicates, from the subgroups categorized by the mass of the satellites, the contribution to the total luminosity within the clusters, the fraction entering the ICL regime, and the contribution to the total ICL luminosity. The filled red, blue, and green circles indicate the subgroups called the ``high-M infaller", ``intermediate-M infaller", and ``low-M infaller", and each error bar cap means the first and third quartiles for each subgroup. The meaning of each subgroup is described in Section \ref{sec:dstate}. The subgroups falling into the clusters of $13.2 < \log M_{\rm{200c}} < 13.6$ are indicated by the light-colored symbols to distinguish them from more massive clusters.}
\label{fig:fstar_z}
\end{figure*}

Recent observational studies have shown the presence of diffuse light in both the group and cluster environments beyond $z=1$ \citep{adami2013,ko2018,demaio2020,coogan2023,joo2023,werner2023}.
Simulation studies also found that the intra-halo stars are ubiquitous in the massive halos, at least since $z=2$ \citep{contini2023,werner2023}.

\begin{figure*}
\centering
\includegraphics[width=0.95\textwidth]{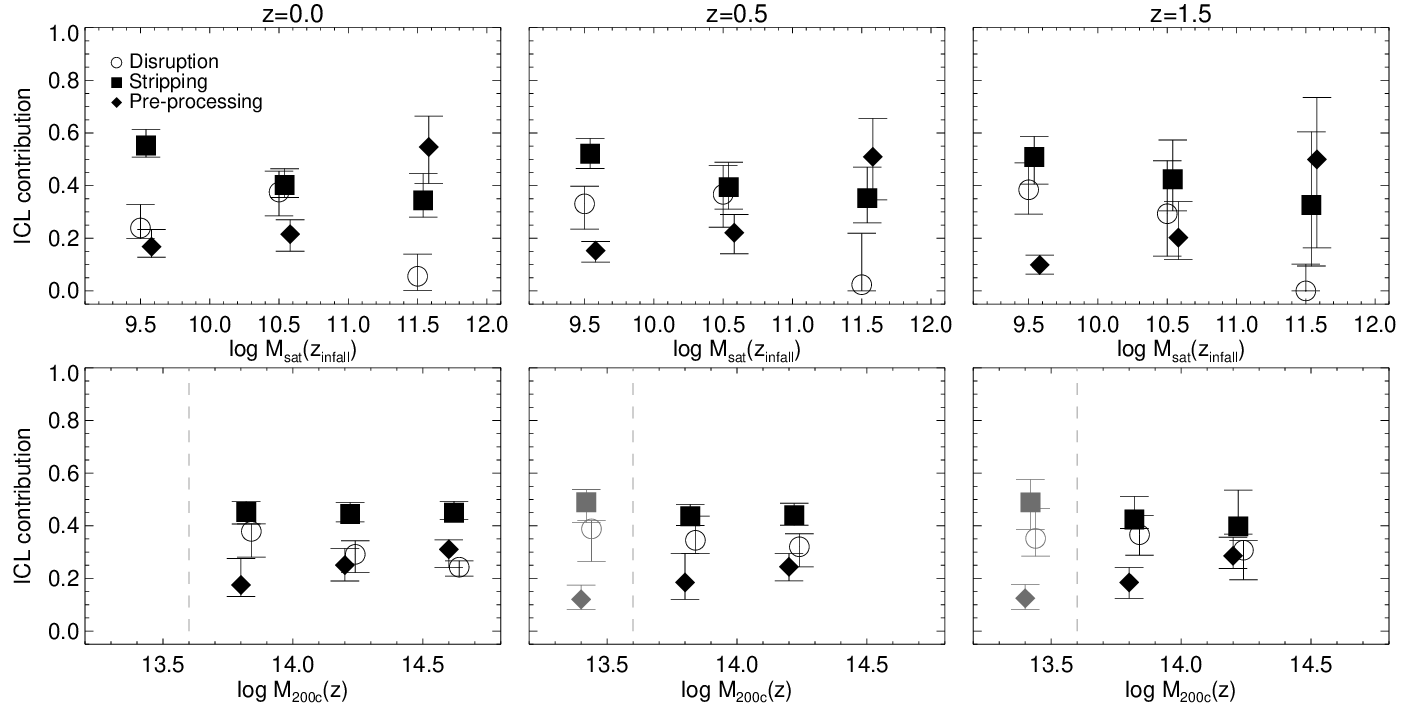}
\caption{The relation between the mass of satellites at $z_{\rm{infall}}$ (upper panels) or the mass of clusters (lower panels) and the contribution of satellites to the ICL within the clusters, categorized depending on the three different formation mechanisms. The open circles, filled squares, and filled diamonds indicate the contribution of disrupted ICL, stripped ICL, and pre-processed ICL to the total ICL stars. From left to right, each panel shows the relation in the clusters at $z=0$, $z=0.5$, and $z=1.5$. The subgroups falling into the clusters of $13.2 < \log M_{\rm{200c}} < 13.6$ are indicated by the light-colored symbols to distinguish them from more massive clusters.}
\label{fig:type_z}
\end{figure*}

In this section, we investigate the contribution of all the satellites accreted onto the host cluster to the ICL stars depending on their galaxy stellar mass at the infall time and the formation mechanisms.
In particular, we focus on the ICL stars at three distinct redshifts, $z=0.0$, $z=0.5$, and $z=1.5$.
Note that we use all snapshot information within a time bin of 500~Myr for each distinct redshift to increase the robustness of our analysis.
As there are no clusters within the most massive host mass bin of $14.4 < \log M_{\rm{200c}} < 14.8$ at $z=0.5$ and $z=1.5$, we extend our investigation to include clusters within the host mass bin of $13.2 < \log M_{\rm{200c}} < 13.6$ at those redshifts. 
This expanded analysis aims to assess how the contribution of the satellites to the ICL stars varies with the mass of host clusters.
We represent them more lightly in Figure \ref{fig:fstar_z} and Figure \ref{fig:type_z} to distinguish them from other massive host halos.
In this section, we do not classify the clusters according to their dynamical state because the classification used in this work is only valid for the clusters at $z=0$, as we mentioned in Section \ref{sec:GRT}.

The top panels of Figure \ref{fig:fstar_z} show the contribution of accreted satellites to the total luminosity of the clusters at three distinct redshifts.
In each panel, we cannot find a noticeable temporal evolution in mixtures of accreted satellites.
In all redshifts, the ``intermediate-M infallers" dominantly contribute to the total luminosity within the clusters at all redshifts.
Moreover, the contribution of the ``low-M infallers" remains relatively consistent ($\sim$20\%) in all clusters, and the contribution of the ``high-M infallers" increases as the mass of clusters increases.
Our results reveal an increase in the contribution of ``high-M infallers" in the clusters with a similar mass over time.
However, it is important to note that this increase is only shown in the host mass bin of $14.0 < \log M_{\rm{200c}} < 14.4$ due to the absence of cluster samples within the most massive host bin at higher redshifts. 

In Section \ref{sec:dstate}, we find that the fraction of the stars in the satellites entering the ICL regime at $z=0$ is unrelated to the dynamical state of clusters (the middle panels of Figure \ref{fig:fstar_z0}).
Therefore, we can expect the same results for all clusters at $z=0$ without distinguishing the dynamical state.
Indeed, in the middle panels of Figure \ref{fig:fstar_z}, we can see while the fraction of ``low-M infallers" transformed to the ICL stars at $z=0$ is as large as 40\%, a smaller fraction of ``intermediate-M infallers" and ``high-M infallers" enters the ICL regime (15-20\%).
Furthermore, the fraction of stars in the satellites entering the ICL regime decreases as the redshift increases.

Even though the ``low-M infallers" do not contribute significantly to the total luminosity of clusters at all redshifts, their importance in the luminosity of ICL stars becomes more apparent as a significant portion of stars in the ``low-M infallers" enters the ICL regime (the bottom panels of Figure \ref{fig:fstar_z}).
They even show the most significant contribution within the host mass bin of $13.2 < \log M_{\rm{200c}} < 13.6$ at $z=0.5$ and $z=1.5$. 
Furthermore, in more massive clusters, their contribution is comparable with the ``intermediate-M infallers".
In contrast, the ``high-M infallers" contribute less to the ICL stars in all redshifts, regardless of the mass of clusters. 
However, there is still a trend that their contribution increases with the increasing mass of clusters in all redshifts.

\begin{figure}
\centering
\includegraphics[width=0.49\textwidth]{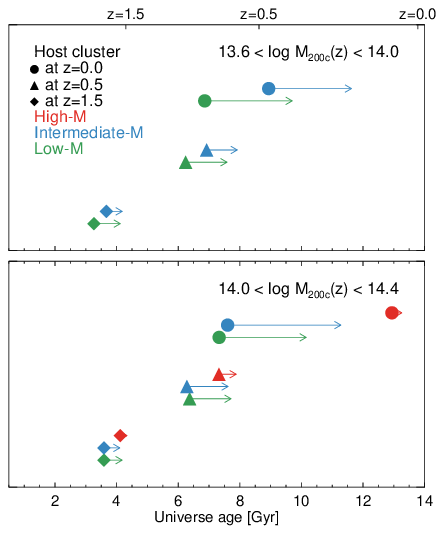}
\caption{The epoch of accretion and transformation of the ICL progenitors in the GRT clusters. The filled symbols and arrowheads indicate when the 80\% of ICL stars at $z=0$ were accreted on the clusters and entered the ICL regime within the clusters. The filled circles, triangles, and diamonds mean the clusters $z=0$, $z=0.5$, and $z=1.5$, and each color of symbols indicates the subgroups categorized by the mass of satellites, shown as Figure \ref{fig:fstar_z0}.}
\label{fig:ficl_z}
\end{figure}

Although the contribution of each infaller to the ICL in the clusters with a similar mass does not show the apparent evolution with the redshift, their accretion history differs depending on the redshift.
Figure \ref{fig:ficl_z} presents when the ICL stars were accreted on the host cluster and transformed into the ICL components within the cluster depending on the redshift.
The filled circles, triangles, and diamonds indicate the time when 80\% of ICL stars at $z=0$, $z=0.5$, and $z=1.5$ were accreted on the host cluster, respectively.
As we mentioned above, there is no cluster within the most massive host bin at $z=0.5$ and $z=1.5$. 
Therefore, we only show the accretion and transformation history of the clusters within the host mass bin of $13.6 < \log M_{\rm{200c}} (z) < 14.0$ (the upper panels) and $14.0 < \log M_{\rm{200c}} (z) < 14.4$ (the lower panels).

This figure shows that, in the clusters at $z=0$, 80\% of ICL stars coming from the ``low-M infallers" and ``intermediate-M infallers" (the filled green and blue symbols) were accreted 5-7~Gyr ago.
It indicates that the stars in these satellites had enough time to enter the ICL regime.
In contrast, this accretion occurred recently at the lifetime of clusters at higher redshift, and thus, the stars in the satellites enter the ICL regime more quickly after they fall into the cluster.
Indeed, each panel shows that the size of arrows of ``low-M infallers" and ``intermediate-M infallers" is shorter in the clusters at higher redshift.
We discuss it further in Section \ref{sec:transformation}.
On the other hand, in the lower panel, we can see that the ``high-M infallers" enter the ICL regime very quickly in the clusters with the host mass bin of $14.0 < \log M_{\rm{200c}} (z) < 14.4$, due to the presence of pre-processed ICL, regardless of the redshift.
We find that more than 50\% of the ICL stars coming from the ``high-M infallers" already entered the ICL regime when their progenitors fell into the cluster within this mass range (the upper panels of Figure \ref{fig:type_z}).
Although, due to the definition of the ICL in this work, the ``low-M infallers" and ``high-M infallers" had the pre-processed ICL when they entered the host cluster, their contribution is the least regardless of the redshift.
Instead, the stripping process within the cluster is the most preferred ICL formation mechanism of these satellites in all redshifts.

As we already see in the bottom panels of Figure \ref{fig:fstar_z}, the ``low-M infallers" and ``intermediate-M infallers" contribute to most of the ICL stars, regardless of the redshift.
Therefore, the stripping process in these satellites is typically the dominant formation mechanism of the ICL in the clusters (the lower panels of Figure \ref{fig:type_z}). 
However, as the contribution of pre-processed ICL increases with the mass of clusters, we note that the pre-processing also can be the main formation mechanism of the ICL in more massive clusters than our GRT clusters as some observed clusters \citep[e.g.,][]{mihos2005,mihos2017,joo2023}.

\subsection{Evolution of the ICL fraction since \lowercase{$z=1.5$}}
\label{sec:evolution2}

\begin{figure*}
\centering
\includegraphics[width=0.95\textwidth]{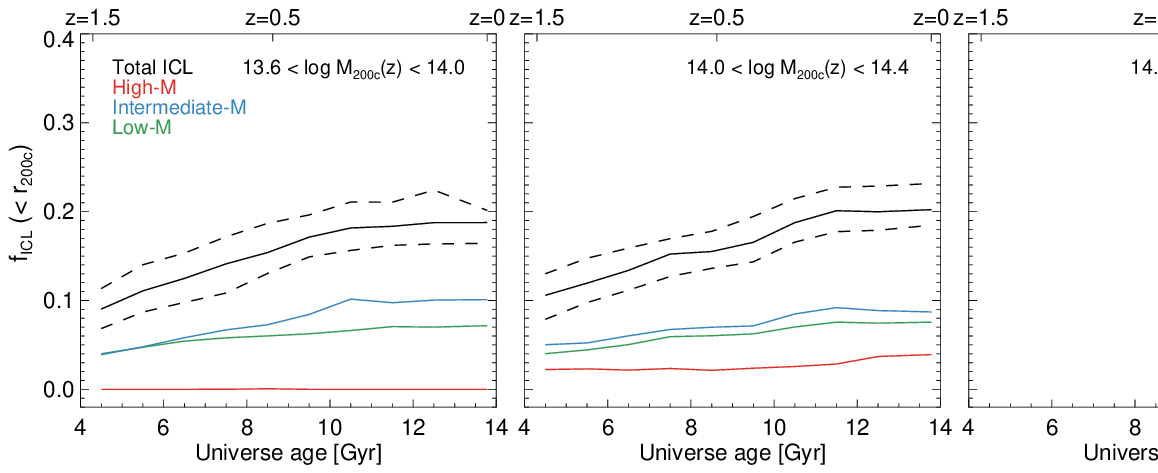}
\caption{Evolution of the ICL fraction within the clusters within the three host mass bins of $13.6 < \log M_{\rm{200c}} < 14$, $14.0 < \log M_{\rm{200c}} < 14.4$, and $14.4 < \log M_{\rm{200c}} < 14.8$ at each redshift. In each panel, the solid and dashed black lines mean the evolution of the median ICL fraction of the clusters and the first and third quartiles of the ICL fraction. The solid red, blue, and green lines indicate the evolution of the ICL fraction contributed from the ``high-M infallers", ``intermediate-M infallers", and ``low-M infallers", respectively.}
\label{fig:fc}
\end{figure*}

Contrary to the unchanged ICL formation mechanism, the ICL fraction evolves with time, even within the clusters of similar mass.
In each panel of Figure \ref{fig:fc}, we measure the ICL fraction of clusters within a fixed mass range at each redshift to eliminate the influence of the host cluster's mass.
Therefore, this figure differs from the right panel of Figure \ref{fig:ficl}, which illustrates the temporal evolution of the ICL fraction in the clusters of fixed mass at $z=0$.
The solid red, blue, and green lines indicate the evolution of the ICL fraction contributed from the ``high-M infallers", ``intermediate-M infallers", and ``low-M infallers", respectively.
These colored lines show that the main contributor to the ICL stars is the ``intermediate-M infallers" at all redshifts, regardless of the mass of clusters.
The solid black line in each panel indicates the median ICL fraction, and the dashed black lines indicate the first and third quartiles of the ICL fraction.
Our results show that the ICL fraction remains constant after $z=0.2-0.3$ (10-11~Gyr), regardless of the mass of clusters.
On the other hand, the ICL fraction in the clusters decreases continuously as the redshift increases.
Despite this decrease, we find that the ICL stars are already significant at the high redshift ($f_{\rm{ICL}}\sim10\%$ at $z=1.5$).
This indicates that significant ICL stars can be formed early in the evolution of the clusters, as already shown in the observation studies \citep[e.g.,][]{ko2018,coogan2023,joo2023}.
Furthermore, the ICL fraction in the clusters at the specific redshift shows insignificant evolution with the mass of clusters, regardless of the redshift, as shown in previous observation and simulation studies \citep[e.g.,][]{zibetti2005,rudick2011,furnell2021,joo2023,ragusa2023}.
However, we pay attention to the fact that the definition of the ICL is different in each study.
In particular, the observation studies have been challenged by the detection limit, as well as have focused on the ICL in more inner regions than $r_{\rm{200c}}$ of the clusters.
To determine the effect of these limits, we investigate changes in the ICL fraction according to the detection limits related to the observable radius and the faint-end SBL for the ICL detection in Section \ref{sec:limit}.

\section{Discussion}
\label{sec:discussion}

\subsection{How long does it take for the stars to enter the ICL regime?} 
\label{sec:transformation}

\begin{figure}
\centering
\includegraphics[width=0.45\textwidth]{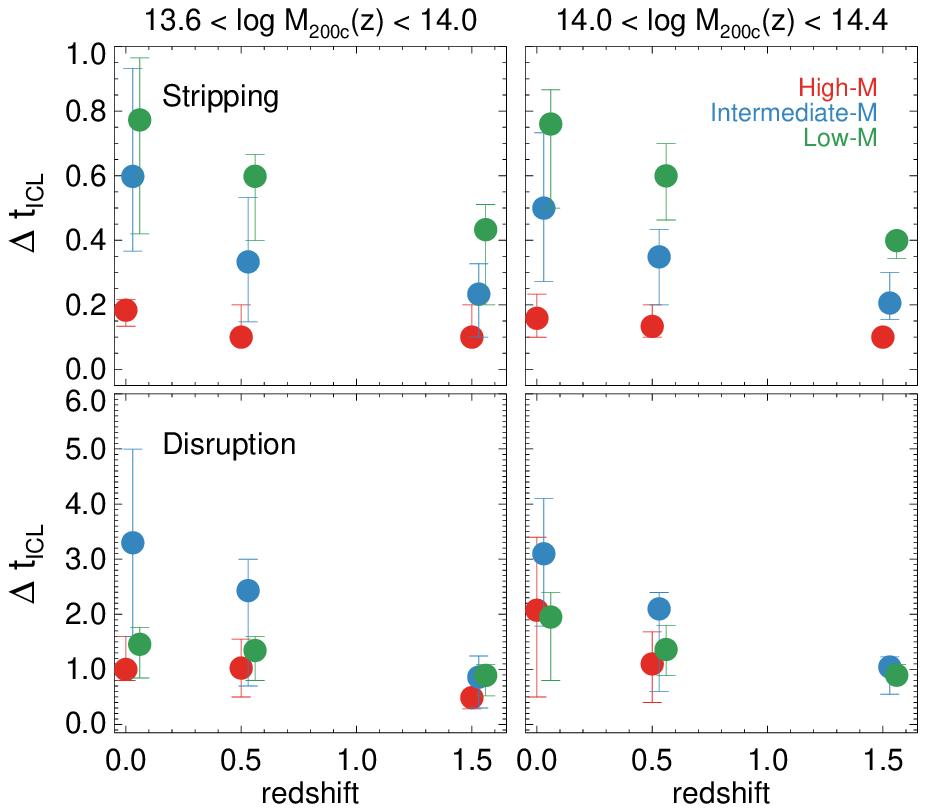}
\caption{Time interval ($\Delta t_{\rm{ICL}}$) required for stars to enter the ICL by the stripping (upper panels) and disruption (lower panels) processes within the clusters at $z=0$, $z=0.5$, and $z=1.5$. The filled red, blue, and green circles indicate the subgroups called the ``high-M infaller", ``intermediate-M infaller", and ``low-M infaller", and each error bar cap means the first and third quartiles for each subgroup. The left and right panels show the relation in the clusters within the three host mass bins of $13.6 < \log M_{\rm{200c}} < 14$, $14.0 < \log M_{\rm{200c}} < 14.4$ at each redshift, respectively.}
\label{fig:dt_icl}
\end{figure}

\cite{chun2023} demonstrated that stars in the low-density outer regions of satellites first contribute to the ICL, and the stripping of stars from the central region is delayed until the outer regions are stripped. 
Consequently, in high-redshift environments where satellites spend a short time after infall, only the outer regions of satellites are expected to enter the ICL regime. 
In contrast, at lower redshifts, where satellites spend enough time within the cluster, the ICL formation through stripping can also occur in the denser central region. 
This is confirmed by the finding that the median density at the infall time of stars contributing to the ICL in clusters at $z=0$ is approximately 1.6 times higher than in clusters with a similar mass at $z=1.5$. 
The upper panels of Figure \ref{fig:dt_icl} show the time interval ($\Delta t_{\rm{ICL}}$) required for individual stars to enter the ICL regime after falling into the host cluster through the stripping process. 
In these panels, $\Delta t_{\rm{ICL}}$ increases as redshift decreases, even when the host clusters have similar masses.

\begin{figure}
\centering
\includegraphics[width=0.45\textwidth]{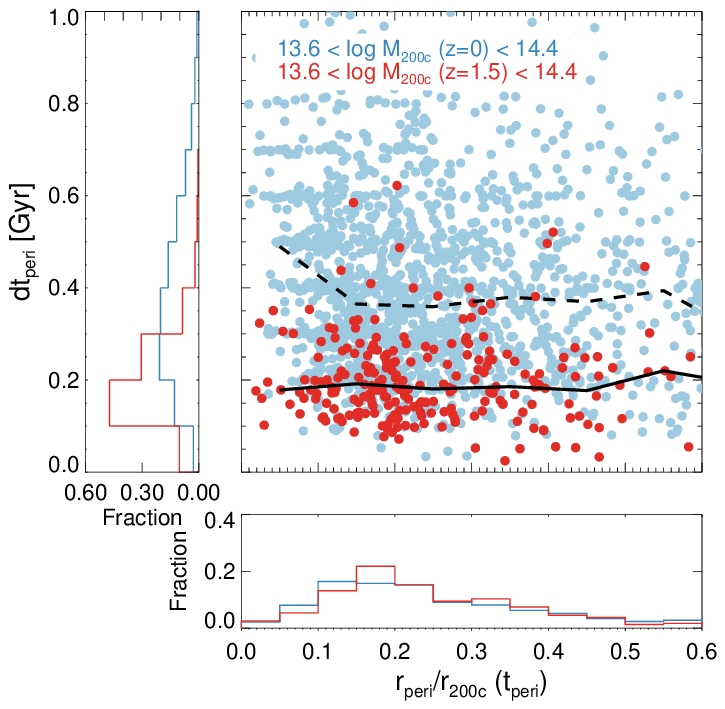}
\caption{The time taken to reach the first pericenter for satellites falling into clusters. A scatter plot indicates the distribution of the position of the first pericenter ($r_{\rm{peri}}/r_{\rm{200c}}$) and the time taken to reach the first pericenter ($dt_{\rm{peri}}$) for satellites. The filled red and blue circles are the satellites falling into the clusters with $13.6 < \log M_{\rm{200c}} < 14.4$ at $z=0$ and $z=1.5$, respectively. The dashed and solid lines mean the median $dt_{\rm{peri}}$ of the satellites at $z=0$ and $z=1.5$ as a function of $r_{\rm{peri}}/r_{\rm{200c}}$. The lower and left histograms show that distributions of $r_{\rm{peri}}/r_{\rm{200c}}$ and $dt_{\rm{peri}}$ depending on the redshift.}
\label{fig:orbit}
\end{figure}

Additionally, satellites experience faster disruption in clusters at higher redshifts (lower panels of Figure \ref{fig:dt_icl}). 
While this may be biased by the shorter staying time of satellites in higher-z clusters, we find that the rapid disruption of satellites at higher redshifts is associated with their smaller orbital periods.
The central panel of Figure \ref{fig:orbit} illustrates the distribution of the first pericenter distance ($r_{\rm{peri}}/r_{\rm{200c}}$) and the time taken to reach the first pericenter ($dt_{\rm{peri}}$) for satellites falling into clusters with $13.6 < \log M_{\rm{200c}} < 14.4$ at $z=0$ (filled blue circles) and $z=1.5$ (filled red circles). 
The dashed and solid black lines represent the median $dt_{\rm{peri}}$ of satellites falling into their host cluster at $z=0$ and $z=1.5$ as a function of the first pericenter distance, respectively. 
The histograms on the left and bottom panels show the distribution of $dt_{\rm{peri}}$ and $r_{\rm{peri}}/r_{\rm{200c}}$ for satellites disrupted before each redshift. 
These results indicate that disrupted satellites in clusters at $z=1.5$ do not pass deeper into the inner regions of the host cluster compared to those in clusters at $z=0$. 
However, satellites in clusters at $z=1.5$ traverse shorter orbital paths to reach the pericenter, owing to the smaller physical size (not comoving scale) of the cluster at the higher redshift. 
As satellites lose a significant portion of their mass after reaching the first pericenter in the cluster \citep[see][]{smith2022a}, it can be anticipated that the satellites in the clusters at $z=1.5$ lose their mass and will be disrupted more quickly, and thus the $\Delta t_{\rm{ICL}}$ by disruption process is shorter as the redshift increases (lower panels of Figure \ref{fig:dt_icl}).

\subsection{Effect of the BCG merger} 
\label{sec:bcg}

Being the most massive galaxy in the cluster, the BCG underwent numerous mergers with other satellites during its assembly. 
In this subsection, we focus on how satellites linked with the growth of the BCG contribute to the ICL within the cluster. 
For this, we investigate how many stars in satellites that experienced the merger with the BCG transform into the ICL regime.
Because we define the satellites and the BCG using the SBL in the projected plane, the spatial overlap between the BCG and satellites may be due to the projection effect.
To alleviate the impact of projection effects, a BCG merger is defined as when more than one-third of satellite stars enter the BCG region in all three distinct projection planes before the disruption of the satellite. 
We classify the satellites experiencing BCG mergers as ``BCG merger infallers" while others are ``non-BCG merger infallers."
It is important to note that the classification of satellites as ``BCG merger infallers" is based on whether they undergo a merger event with the BCG before their disruption. 
Consequently, the contribution of ``BCG merger infallers" to the ICL encompasses stars that enter the ICL regime through processes such as stripping and pre-processing before these satellites interact with the BCG.

\begin{figure*}
\centering
\includegraphics[width=0.95\textwidth]{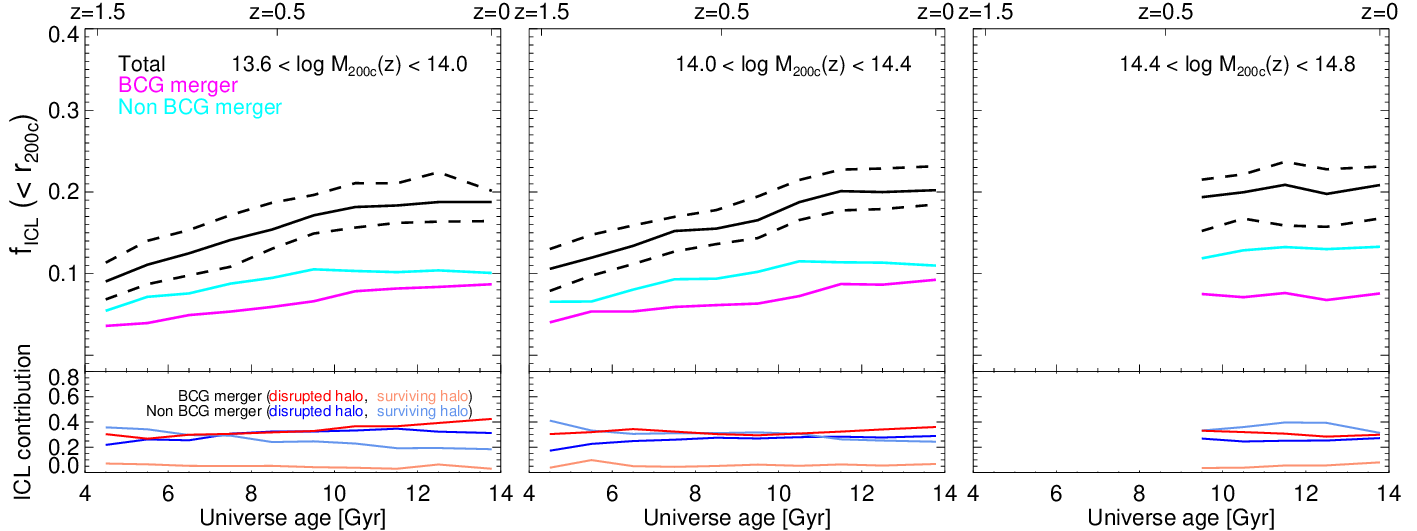}
\caption{Evolution of the ICL fraction of the clusters within the three host mass bins of $13.6 < \log M_{\rm{200c}} < 14$, $14.0 < \log M_{\rm{200c}} < 14.4$, and $14.4 < \log M_{\rm{200c}} < 14.8$ at each redshift. In each upper panel, the solid and dashed black lines mean the evolution of the median ICL fraction of the clusters and the first and third quartiles of the ICL fraction. The solid magenta and cyan lines indicate the evolution of the ICL fraction contributed from the ``BCG merger infallers" and ``non-BCG merger infallers", respectively. The lower panels show the contribution of the ``BCG merger infallers" and ``non-BCG merger infallers" to the total ICL within the clusters. In these panels, the ``BCG merger infallers" (or ``non-BCG merger infallers") are divided by disrupted satellites before each redshift, represented by the deep red (or blue) line, and the surviving satellites, represented by the light red (or blue) line.}
\label{fig:fc_mg}
\end{figure*}

Figure \ref{fig:fc_mg} shows the time evolution of the ICL fraction depending on the mass of clusters.
The upper panels display the evolution of the ICL fraction in clusters with a similar mass, with the solid black line indicating the median ICL fraction and dashed lines denoting the first and third quartiles. 
The solid magenta and cyan lines represent the contributions of ``BCG merger infallers" and ``non-BCG merger infallers," respectively. 
Across the mass of host clusters, the contribution of ``non-BCG merger infallers" consistently outweighs that of ``BCG merger infallers." 
This trend persists across redshifts. 
However, the lower panels reveal that the importance of ``non-BCG merger infallers" is retained by different types of satellites depending on the redshift.
In these panels, contributions from ``non-BCG merger infallers" are further divided into those from disrupted satellites (the solid deep blue line) before each redshift and surviving satellites (the solid light blue line). 
Notably, surviving satellites contribute nearly twice as much as disrupted satellites to the ICL components of ``non-BCG merger infallers" at $z=1.5$. 
While the contribution of disrupted satellites to the ICL rises due to continuous disruptions of satellites within the cluster over time, the significance of surviving satellites becomes relatively more pronounced with higher cluster masses, especially at lower redshifts. 
This trend is attributed to the increased survivability of satellites in more massive clusters, as supported by studies such as \cite{bahe2019}. 

Although our results show the relative importance of the ``non-BCG merger infallers" to the formation of the ICL, the contribution of the ``BCG merger infallers" is also significant (40-45\%).
In particular, most of the ``BCG merger infallers" are disrupted (the solid deep red line) before each redshift.
It is consistent with recent studies indicating that ICL stars associated with the growth of the BCG are significant \citep[e.g.,][]{jimenez-teja2023,joo2023}.
We especially find that their importance increases when focused only on the inner regions of clusters.
We discuss it further in Section \ref{sec:limit}.

\subsection{The effect of the detection limits on the ICL stars} 
\label{sec:limit}

Previous studies have highlighted the recoverability of faint low surface brightness structures through the application of lower faint-end SBL that enable deeper observation \citep[e.g.,][]{burke2015,mihos2017,martin2022}. 
\cite{burke2015}, in particular, emphasized the necessity of quantifying the impact of the faint-end SBL on the ICL fraction due to its strong dependence on this limit.
In this section, we examine the influence of detection limits on ICL stars, with a focus on clusters within $14.0 < \log M_{\rm{200c}} < 14.4$. 
It is noteworthy that group-sized halos ($13.6 < \log M_{\rm{200c}} < 14.0$) exhibit similar results, although we do not show the results.

\begin{figure}
\centering
\includegraphics[width=0.45\textwidth]{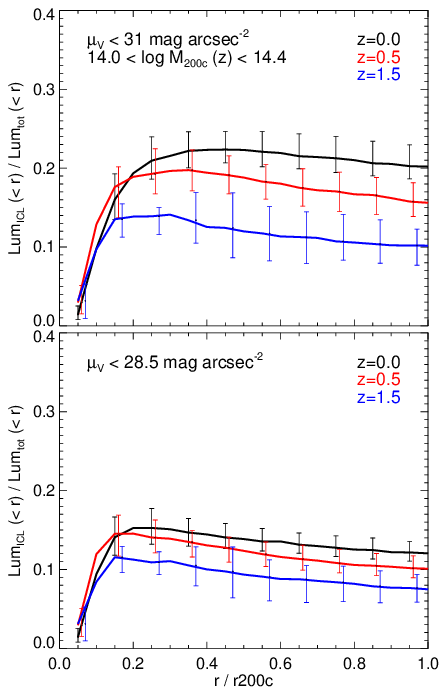}
\caption{Cumulative luminosity fraction of ICL stars as a function of distance from the BCG in the clusters of $14.0 < \log M_{\rm{200c}} < 14.4$. The distance is normalized by $r_{\rm{200c}}$. The V-band faint-end surface brightness is considered down to $\mu_{V}=31~\rm{mag}~\rm{arcsec^{-2}}$ in the upper panel and $\mu_{V}=28.5~\rm{mag}~\rm{arcsec^{-2}}$ in the lower panel. In each panel, the solid black, deep red, and deep blue lines indicate the radial profiles of the ICL fractions in the clusters at $z=0$, $z=0.5$, and $z=1.5$, respectively. Each error bar cap indicates the first and third quartiles of the ICL fractions at each radius.}
\label{fig:profile}
\end{figure}

Figure \ref{fig:profile} illustrates the cumulative luminosity fraction of ICL stars within clusters of $14.0 < \log M_{\rm{200c}} < 14.4$ as a function of distance from the BCG, normalized by $r_{\rm{200c}}$. 
The V-band faint-end surface brightness is considered down to $\mu_{V}=31~\rm{mag}~\rm{arcsec^{-2}}$ in the upper panel and $\mu_{V}=28.5~\rm{mag}~\rm{arcsec^{-2}}$ in the lower panel. 
In the upper panel, the ICL fraction within $r_{\rm{200c}}$ of the clusters decreases as redshift increases, but the ICL is already abundant ($\sim$10\%) at $z=1.5$. 
However, the lower panel demonstrates that applying a higher faint-end SBL leads to an underestimation of the ICL fraction, with this underestimation becoming more pronounced as redshift decreases.
Specifically, compared to tracing ICL stars with $\mu_{V} < 31~\rm{mag}~\rm{arcsec^{-2}}$, the ICL fraction within $r_{\rm{200c}}$ decreases by around 25\% at $z=1.5$ and about 40\% at $z=0.0$. 
These substantial decreases highlight that we miss numerous ICL stars within $r_{\rm{200c}}$ due to the detection limit.

While the two panels indicate the changes in ICL fraction by choice of faint-end SBL, both also reveal that the ICL fraction increases when focusing further inside due to the decreasing contribution of surviving satellites to the total stellar luminosity within the clusters. 
However, the presence of the BCG stars significantly reduces the ICL fraction in the central region. 
This highlights, for accurate comparisons between studies on ICL fraction, the need to carefully consider the effects of detection limits related to the observable radius and the faint-end SBL.

\begin{figure*}
\centering
\includegraphics[width=0.95\textwidth]{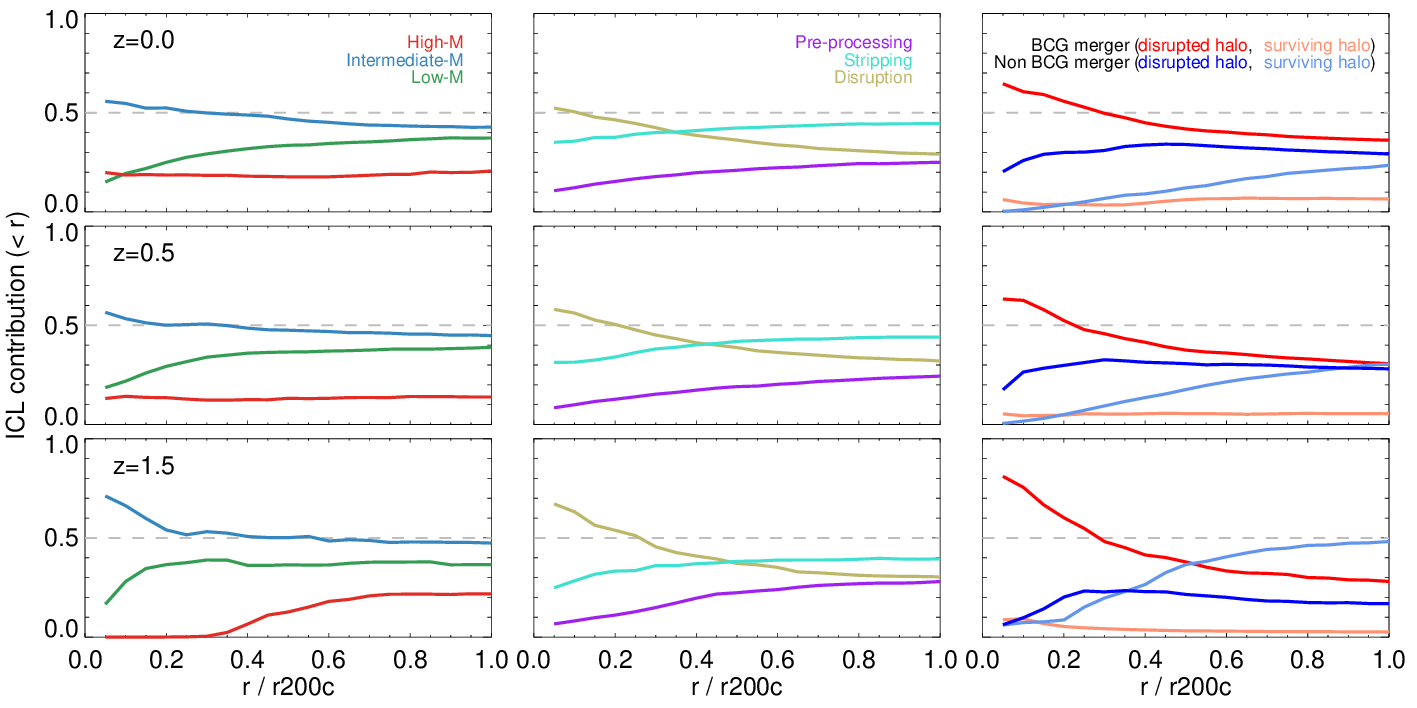}
\caption{Cumulative luminosity fraction of ICL stars as a function of distance from the BCG in the clusters of $14.0 < \log M_{\rm{200c}} < 14.4$ depending on the formation channels. The distance is normalized by $r_{\rm{200c}}$. Each column of panels, starting from the left, represents the contribution of the subgroups categorized by satellite mass, formation mechanisms, and their relation with the BCG to the ICL at $z=0$, $z=0.5$, and $z=1.5$. See the text for a description of the subgroups.}
\label{fig:profile2}
\end{figure*}

As the detection limits can influence the ICL fraction of the clusters, there is a possibility that they also affect the main formation channel. 
Figure \ref{fig:profile2} illustrates the cumulative ICL contribution as a function of distance from the BCG, normalized by $r_{\rm{200c}}$. 
Each column of panels, starting from the left, represents how satellites contribute to the ICL at $z=0$, $z=0.5$, and $z=1.5$, depending on satellite mass, formation mechanisms, and their relation with the BCG. 
The analysis focuses on all ICL stars ($\mu_{V} < 31~\rm{mag}~\rm{arcsec^{-2}}$) within clusters, but we find that the choice of the faint-end SBL does not change the results.

In the left panels, it is evident that the ``intermediate-M infallers" (the solid blue line) consistently contribute as the main formation channel for ICL stars within $r_{\rm{200c}}$ of the clusters, regardless of redshift, as shown in Section \ref{sec:evolution1} (see Figure \ref{fig:fstar_z}).
Their contribution to the ICL increases when focusing on the inner region. 
Conversely, the ``high-M infallers" (the solid red line) maintain a constant contribution to the ICL irrespective of the distance from the BCG, except at $z=1.5$.
At $z=1.5$, they cannot reach the central region, resulting in a decrease in their ICL contribution towards the inner region. 
This decreasing contribution is also observed for ``low-M infallers" (the solid green line) due to their preference for contributing to the ICL rather than the BCG.

The middle panels reveal that disrupted ICL dominates near the BCG, where the strong tidal field more efficiently disrupts satellites (the solid khaki line) at all redshifts. 
Conversely, pre-processed (the solid purple line) and stripped ICL (the solid turquoise line) stars contribute more with increasing distance from the BCG. 
In particular, the stripping process dominates the formation of ICL stars within $r_{\rm{200c}}$ of the clusters, as shown in Figure \ref{fig:type_z}. 

The right panels demonstrate, as the satellites stay enough time within the host cluster at low redshift, the contribution of the disrupted satellites to the ICL within $r_{\rm{200c}}$ of the clusters is significantly larger than that of the surviving satellites.
On the other hand, as the redshift increases, the contribution of surviving satellites increases, especially for the surviving satellites (the solid light blue line) of the ``non-BCG merger infallers", reaching up to 50\%. 
Although the ICL stars coming from them spread throughout the host cluster, we should note that some ICL stars can still be observed around the surviving parent galaxies or groups and can significantly grow the ICL fraction \citep[e.g.,][]{mihos2017,chun2023}.
The panels also show that the ``BCG merger infallers" contribute a substantial portion (40-45\%) of ICL stars within $r_{\rm{200c}}$ of the clusters (see also Figure \ref{fig:fc_mg}), with their contribution increasing as focusing the inner region.
In particular, the disrupted satellites (the solid deep blue line) of the ``BCG merger infallers" contribute more than half of the ICL stars within 0.3-0.4$r_{\rm{200c}}$ of the clusters, regardless of the redshift.
The fact that numerous ICL stars are associated with the merger tree of the BCG is in agreement with the previous studies \citep[e.g.,][]{murante2007,harris2017}.

These results show that the importance of ICL formation channels is influenced by the detection limit, with even the dominant channel subject to change.
Moreover, while we distinguish the ICL and the BCG using the SBL in the surface brightness map, their separation can also be achieved using different decomposition methods that potentially yield different ICL fractions. 
\citet{kluge2021} investigated various ways to separate the ICL and the BCG in 170 low-redshift ($z < 0.08$) clusters. 
They found a large scatter in the fraction of ICL luminosity relative to the total luminosity of the ICL and the BCG, ranging from 34\% to 71\%, when employing four different decomposition methods.
To test the effect of the methodology on the determination of ICL components, we additionally measure the ICL fraction using the double S\'ersic decomposition (DS) method\footnote{In this test, we use 10 clusters at $z=0$, $z=0.5$, and $z=1.5$, respectively. These clusters have lower $\chi^2$ values for the double S\'ersic fit than other clusters at each redshift.}.
For this, we restrict our measurement of the ICL to within $0.3R_{\rm{200c}}$ of the host cluster. This is necessary because most of the ICL stars beyond this radius are located on the outskirts of surviving satellites, and they may interfere with the fitting process.

Our analysis reveals that the ICL fraction ($< 0.3R_{\rm{200c}}$) derived from the SBL method is underestimated by $\sim$12 percentage points compared to those obtained from the DS method. 
This underestimated ICL fraction in the SBL method has also been reported in previous studies \citep[e.g.,][]{cooper2015,montes2018,kluge2021}.
Although there is no significant change in the importance of ICL formation channels between the two distinct methods, we find that the contribution of ``high-M infallers” to the ICL in the DS method generally increases compared to the SBL method, while the contribution of ``low-M infallers” decreases. 
This is because the ICL components obtained from the DS method include more stars in the inner region where the contribution of ``high-M infallers” increases compared to those obtained from the SBL method (see the left panels of \ref{fig:profile2}). 
Our findings emphasize the need for not only accurate quantification of detection limits about the ICL but also caution when comparing with observational studies that use a different approach.

\section{Summary}
\label{sec:summary}

Using cosmological N-body simulations, we investigate the evolution of the ICL in the 84 clusters of $13.6 < \log M_{\rm{200c}} < 14.4$ and measure their ICL fraction from $z=1.5$ to $z=0.0$.
To describe the formation and evolution of the stellar structures in the cosmological N-body simulations, we adopt an alternative simulation technique referred to as the ``galaxy replacement technique" (GRT), first introduced in \cite{chun2022}.
As GRT does not include computationally expensive baryonic physics, we accurately trace the spatial distribution and evolution of diffuse stellar structures in clusters with sufficient high-resolution star particles.

In this study, we aim to understand how the formation channels of the ICL differ depending on the dynamical state of the clusters at $z=0$ or as a function of redshift.
Moreover, we investigate the effect of the detection limits on the ICL properties.
We summarize our results as follows:

\begin{enumerate}\setlength{\itemsep}{-1mm}
\item The accretion of ``low-M infallers" and ``intermediate-M infallers" onto relaxed clusters precedes that onto unrelaxed clusters due to the earlier growth of relaxed clusters. 
This results in the ICL fraction of relaxed clusters being higher, as the satellites have more time to enter the ICL regime.
\item The ICL fraction of the clusters with a similar mass decreases as the redshift increases, but the ICL is already abundant ($\sim$10\%) at $z=1.5$.
However, there is no significant evolution of the ICL fraction with the mass of clusters.
\item At all radii, the typical progenitors of the ICL fall into all clusters with the ``intermediate-M infallers", regardless of the redshift and properties of the clusters.
However, the contribution of the ``high-M infallers" tends to be more important in dynamically unrelaxed clusters (at $z=0.0$) or more massive clusters.
\item Regardless of the redshift and properties of the clusters, the stripping process of the surviving satellites is the dominant formation mechanism of the ICL within the $r_{\rm{200c}}$ of the clusters, but the stars from the disrupted satellites dominantly contribute to the ICL in an inner region.
On the other hand, the contribution of pre-processed ICL stars, primarily from ``high-M infallers", tends to be related to the properties of the clusters, i.e., their contribution is more important in dynamically unrelaxed clusters (at $z=0.0$) or more massive clusters.
\item At all redshifts, the numerous ICL stars (40-45\%) within $r_{\rm{200c}}$ of the clusters are associated with the merger tree of the BCG.
The disrupted satellites among the ``BCG merger infallers" even contribute more than half of the ICL stars within $0.3-0.4r_{\rm{200c}}$ of the clusters.
\end{enumerate}

These results are derived using all stellar components brighter than $\mu_{V} = 31~\rm{mag}~\rm{arcsec^{-2}}$ corresponding to the resolution limit of the GRT simulation.
However, we also find that the detection limits related to the observable radius and the faint-end SBL hide numerous ICL stars and even change the dominant formation channel of the ICL.
Therefore, our results emphasize the need for accurate quantification of detection limits to compare the studies about the ICL.
We expect that the presence of faint ICL stars missed by the detection limits can be revealed by upcoming deeper imaging observations such as Legacy Survey of Space and Time (LSST), Euclid, K-DRIFT, etc.



\bibliography{main}{}
\bibliographystyle{aasjournal}

\end{document}